\documentclass[aps,amsmath,amssymb,prl,twocolumn,preprintnumbers,superscriptaddress,nobalancelastpage]{revtex4}
\usepackage{graphicx}
\newcommand{\be}{\begin{eqnarray}}
\newcommand{\ee}{\end{eqnarray}}

\begin{document}
\title{Quantum Belief Propagation}
\author{M.~B.~Hastings}
\affiliation{Center for Nonlinear Studies and Theoretical Division,
Los Alamos National Laboratory, Los Alamos, NM 87545, hastings@lanl.gov}
\begin{abstract}
We present an accurate numerical algorithm, called quantum belief propagation
(QBP), for simulation of one-dimensional quantum systems at non-zero
temperature.  The algorithm exploits the fact that quantum effects
are short-range in these systems at non-zero temperature, decaying
on a length scale inversely proportional to the temperature.  We compare
to exact results on a spin-$1/2$ Heisenberg chain.  Even a
very modest
calculation, requiring diagonalizing only $10$-by-$10$ matrices,
reproduces the peak susceptibility with a relative
error of less than 
$10^{-5}$, while more elaborate calculations further reduce the error.
\end{abstract}
\maketitle

The fact that interactions are short-ranged in many physical systems
is a major simplification in finding quantum ground states.
The classic example of
this is the density matrix renormalization group(DMRG)\cite{dmrg}, which
relies on the ability to approximate the ground state by a matrix product
state.  While it has long been believed that such an approximation is
possible whenever there is a spectral gap, due to conformal field theory
calculations\cite{cft} and decay of correlations\cite{loc1,loc2},
only very recently has a general proof been given that such an approximation
is possible whenever there is a local Hamiltonian and a gap\cite{areal}.

At non-zero temperature, the system is in a
mixed state instead.  Here, matrix product density
operators\cite{mpdo} play the same role that
matrix product states do in studying pure states,
and a non-zero temperature plays a similar role
when it comes to representing mixed states as matrix product
density operators as does a gap for representing pure states as matrix product
states.
Indeed, it has been shown that good
matrix product operator representations exist for quantum systems at
any non-zero temperature\cite{thermal}.  In this paper we present
quantum belief propagation (QBP), another method
for finding matrix product density operators for
thermal states, which avoids the problem of Trotter error in other
methods.

Belief propagation\cite{bp}
for classical systems (CBP) is essentially a Bethe-Peierls solution
of a classical statistical mechanics model.  CBP is exact on trees,
and is often a very good approximation on lattices with few loops.  At
the end of this paper we will discuss application of the QBP equations
to higher dimensional systems and trees, but for now we will focus discussion
on belief propagation for one dimensional lattices.  In this case, CBP
become equivalent to a transfer matrix technique: one solves the problem on
a chain of $N$-sites to get a partition function which depends on the value
of the spin on the $N$-th site.  Then, one adds a coupling of the $N$-th
spin to one additional spin, traces out the $N$-th spin, arriving at a partition
function which depends on the value of the spin on the $N+1$-st site.
One proceeds in this way, iteratively solving longer and longer chains.

In a quantum system, we will proceed in a similar way.  However, now the
operator coupling the $N$-th spin to the $N+1$-st spin need not commute with
the rest of the Hamiltonian.  This means that to study properties of the
chain of $N+1$-spins, it is not sufficient simply to know the density matrix
of the $N$-th spin in an $N$-site chain.  However, our physical intuition
tells us that at a non-zero temperature, quantum effects should be
short range.  Our main result in the next section realizes this intuition
in the QBP
equations, which involve two terms.  One is a ``classical" term which is local,
coupling the $N$-th spin to the $N+1$-st spin.  The other is a ``quantum" term,
which is non-local, coupling the $N+1$-st spin to several other spins; however,
this quantum term is exponentially decaying on a length scale set by
the inverse temperature.  This will allow us to accurately describe statistical
properties of the $N+1$-st spin knowing only the reduced density matrix
of spins $N,N-1,...,N-l_0+1$, for some $l_0$, so that we keep track
of a reduced density matrix for $l_0$ spins.  We then iterate
these equations by tracing out spin $N-l_0+1$ and then computing
statistical properties of spin $N+2$, keeping always a reduced density
matrix on the last $l_0$ spins on the chain.

{\it Quantum Belief Propagation Equations---}
The QBP equations describe how the
partition function, $\exp(-\beta H)$, changes when $H$ is changed by
some perturbation $A$.  
We take $A,H$ Hermitian throughout.
We will apply this result to
the following case: we have a nearest-neighbor Hamiltonian,
with $h_{i,i+1}$ the term coupling spin $i$ to spin $i+1$.  We will
take $H=h^{(N)}
\equiv \sum_{i=1}^{N-1} h_{i,i+1}$, so that $H$ is the Hamiltonian for
the first $N$ spins, and $A=h_{N,N+1}$.
We define
\be
H_s=H+s A.
\ee
Then, $H_1=h^{(N)}+A=h^{(N+1)}$.

Although ultimately we want to compute $\exp(-\beta H_s)$ at $s=1$, we
begin by computing the derivative $\partial_{s}
\exp(-\beta H_s)$.
Let $A_{ab}(s)$ denote
matrix elements of $A$ in a basis of eigenstates of $H_s$, with energies
$E_a(s),E_b(s)$.  Define the operator $A^{\omega,s}$ by its matrix elements:
$(A^{\omega,s})_{ab}=A_{ab}(s) \delta(E_a(s)-E_b(s)-\omega)$.  We will
do most of the calculations in terms of $A^{\omega,s}$ rather than
$A$ for simplicity,
although we will convert certain results back into results
in terms of $A$ itself using the integral
\be
\label{qbp1}
\Bigl(\partial_{s} \exp(-\beta H_s)\Bigr)=
\int {\rm d}\omega 
\Bigl(\partial_{\epsilon} \exp[-\beta (H_s+\epsilon A^{\omega,s})]\Bigr),
\ee
where all derivatives with respect to $\epsilon$ are taken at
$\epsilon=0$ throughout.

One result for the derivative on the right-hand side
of Eq.~(\ref{qbp1}) (not the result we will use!) is
$\label{beforeafter}
\Bigl(\partial_{\epsilon} \exp[-\beta(H_s+\epsilon A^{\omega,s})]\Bigr)
=-\exp(-\beta H_s) 
\Bigl( \frac{\exp(\beta \omega)-1}{\omega}
A^{\omega,s} \Bigr)$.
One problem with this is that the operator norm
$\Vert\frac{\exp(\beta \omega)-1}{\omega}
A^{\omega,s}\Vert$ may be exponentially large.
The QBP equations will
improve on this, writing $\partial_s\exp(-\beta H_s)=\eta\exp(-\beta H_s)
+\exp(-\beta H_s)\eta^{\dagger}$, where
$\Vert \eta_s \Vert
\leq (\beta/2) \Vert A \Vert$.  
Finally, $\eta$ will be a local operator as discussed below\cite{locEXP}.

To find the QBP equations, we begin by
studying a certain correlation function.  Let $B$ be an arbitrary
operator.  Then,
\begin{eqnarray}
\label{corrln1}
&&\partial_{\epsilon} {\rm tr}(\exp[-\beta(H_s+\epsilon A^{\omega,s})] B)
\\ \nonumber & = &
-\int_0^{\beta} {\rm d} \tau
{\rm tr}(\exp(-\beta H_s) A^{\omega,s}(-i\tau,H_s) B) \\ \nonumber
& = &
-\frac{\exp(\beta \omega)-1}{\omega}
{\rm tr}(\exp(-\beta H_s) A^{\omega,s} B).
\end{eqnarray}
Adding and subtracting $-(\beta/2) {\rm tr}(\{\exp(-\beta H_s),
A^{\omega,s}\} B)$ to Eq.~(\ref{corrln1}) gives
\begin{eqnarray}
\label{corrln2}
&&\partial_{\epsilon} {\rm tr}(\exp[-\beta(H_s+\epsilon A^{\omega,s})] B)
\\ \nonumber
&=&
-\frac{\beta}{2} {\rm tr}(\{\exp(-\beta H_s),A^{\omega,s}\} B)
\\ \nonumber
&+&\Bigl(
\frac{\beta}{2} (1+e^{\beta \omega})
-\frac{e^{\beta \omega}-1}{\omega}
\Bigr)
{\rm tr}(\exp(-\beta H_s) A^{\omega,s} B)
\end{eqnarray}
where we used
${\rm tr}(A^{\omega,s} \exp(-\beta H_s) B)= \exp(\beta \omega)
{\rm tr}(\exp(-\beta H_s)A^{\omega,s} B)$.

We now focus on the correlation function
${\rm tr}(\exp(-\beta H_s) A^{\omega,s} B)$, following
a procedure very similar to that in \cite{fermi}.
In \cite{fermi} the result was expressed in
terms of anti-commutators, as $A,B$ were fermionic operators and hence
anti-commutated if they were separated in space, while here we will
express the result in terms of commutators, since we intend to apply it
to bosonic operators where the Lieb-Robinson bound is expressed as a
bound on the commutator.  Using 
${\rm tr}(\exp(-\beta H_s)A^{\omega,s} B)=
\frac{1}{1-\exp(\beta \omega)}
{\rm tr}(\exp(-\beta H_s)[A^{\omega,s},B])$,
we have
\begin{eqnarray}
&&\Bigl(
\frac{\beta}{2} (1+e^{\beta \omega}) 
-\frac{e^{\beta \omega}-1}{\omega}
\Bigr)
{\rm tr}(\exp(-\beta H_s)A^{\omega,s} B) \nonumber \\ &=&
\beta F(\omega) 
{\rm tr}(\exp(-\beta H_s)[A^{\omega,s},B]),
\end{eqnarray}
where
\be
F(\omega)\equiv
\frac{1}{2}
\frac{1+e^{\beta \omega}}{1-e^{\beta\omega}}
+\frac{1}{\beta \omega}
=
-\frac{\coth(\beta \omega/2)}{2}+\frac{1}{\beta \omega}.
\ee
Thus for any operator $B$, we have
$\partial_{\epsilon} {\rm tr}(\exp[-\beta(H_s+\epsilon A^{\omega,s})] B)=
-\frac{\beta}{2} {\rm tr}(\{\exp(-\beta H_s),A^{\omega,s}\} B)
+\beta F(\omega) 
{\rm tr}(\exp(-\beta H_s)[A^{\omega,s},B])$,
and hence
\begin{eqnarray}
\label{clq}
&&\partial_{\epsilon} \exp[-\beta(H_s+\epsilon A^{\omega,s})] \\ \nonumber & = &
-\frac{\beta}{2}\{\exp(-\beta H_s),A^{\omega,s}\}
+\beta F(\omega)[\exp(-\beta H_s),A^{\omega,s}].
\end{eqnarray}
We define
\begin{eqnarray}
\eta^\omega_s&=&-(\frac{\beta}{2}+\beta F(\omega)) A^{\omega,s}, \\ \nonumber
\eta_s& =& \int {\rm d}\omega \eta^\omega_s
\end{eqnarray}
and integrate Eq.~(\ref{clq}) over $\omega$ to get
\be
\label{clqI}
\partial_s \exp(-\beta H_s)=
\eta_s\exp(-\beta H_s)+\exp(-\beta H_s) \eta_s^{\dagger}.
\ee

Eq.~(\ref{clqI}) is the QBP equation for
$\partial_s \exp(-\beta H_s)$.
The anti-commutator in
Eq.~(\ref{clq}) is a ``classical term".  It is the only
term present if $[A,H]=0$, in which case these equation reproduce the
classical belief propagation equations.
The commutator in Eq.~(\ref{clq}) is a ``quantum"
term.  It is odd in $\omega$ and vanishes at $\omega=0$.
We now discuss locality properties of the
quantum term, assuming that $H$ is local in the sense of having
a Lieb-Robinson bound\cite{lr}.  
We have\cite{fermi}
$\beta\int {\rm d}\omega F(\omega)A^{\omega,s}=$
\begin{eqnarray}
&&\int {\rm d}\omega 
\Bigl(-\frac{\beta}{2} \coth(\beta \omega/2)+\frac{1}{\omega} \Bigr)
A^{\omega,s}
\\ \nonumber
&=& 
-\int {\rm d}\omega 
\sum_{n\neq 0} (\omega-2\pi n i/\beta)^{-1}
A^{\omega,s} \\ \nonumber
&=& i\sum_{n\geq 1}\int_{-\infty}^{\infty} {\rm d}t \; {\rm sign}(t) 
\exp(-2\pi n t/\beta)
A(t,H_s),
\end{eqnarray}
where the sum is over integer $n$
and $A(t,H_s)=\exp(i H_s t) A \exp(-i H_s t)$.
The integral over $t$ is exponentially
decaying for $t \gtrsim \beta$ and so, using a Lieb-Robinson bound,
$\eta$ is local.

{\it Numerical Implementation and Results---}
In this section we discuss the numerical implementation of the
QBP equations, and the results of their application to the antiferromagnetic
spin-$1/2$ Heisenberg chain.  The idea is to take the QBP
equations, which depend on $A$ and $H=h^{(N)}$, and instead set
$H=\sum^{N-1}_{i=N-l_0+2} h_{i,i+1}$, for some constant $l_0\geq 2$.
Since $\eta$
is local, this approximation is justified for small enough $\beta$; as
$\beta$ increases, $l_0$ must increase and the numerical effort
is exponential in $l_0$, since one must diagonalize
matrices of size $2^{l_0}$.

Although translation invariance is {\it not} necessary,
it is useful as
we can then apply translations to
$H=\sum^{N-1}_{i=N-l_0+2} H_{i,i+1}$ to make $H$ equal to $h^{(l_0-1)}$,
so for the Heisenberg chain with coupling constant $J=1$, 
$H=\sum_{i,1\leq i \leq l_0-2} \vec S_i \cdot \vec S_{i+1}$.  We
set $A=h_{l_0-1,l_0}$.
We set $T$ to be the operator which translates
one site to the right.
We define $\rho$ to be a reduced density matrix for sites $1...l_0$, so
that $\rho$ is a $2^{l_0}$ dimensional matrix.
Define
\be
\label{odefn}
O={\cal S}' \exp(\int_0^1 \eta_{s'} {\rm d}s'),
\ee
where the ${\cal S}'$ denotes that the integral is $s'$-ordered.

The algorithm to compute the free energy per site of an infinite chain
proceeds through the following four steps.  $(1)$ Initialize
$\rho$ to $\exp(-\beta H)$.  $(2)$ Approximately compute the operator
$O$ as described below.  $(3)$ Go through a series of $n_{it}$ iterations
of the following three steps: 

(A) replace $\rho$ with $O \rho O^{\dagger}$.
(B) trace out the first site, so that $\rho$ is replaced with
${\rm tr}_1(\rho)\otimes \openone_{l_0+1}$.  Here, ${\rm tr}_1$ denotes
a partial trace
over the first
site, and $\openone_{l_0+1}$ is the unit operator on the $l_0+1$-th site.  At
this stage $\rho$ is now a density operator on sites $2...l_0+1$.  Translate
by one site, so that $\rho$ becomes a density operator on sites $1...l_0$
again.
(C) Define $Z={\rm tr}(\rho)$ and then replace $\rho$ with
$Z^{-1} \rho$.

After these iterations, we (4) Output $\ln(Z)$ from the last step as the
free energy.  This procedure relies on a series of iterations of
steps $(A),(B),(C)$ to find the $\rho$ which is the fixed point
of the map
\be
\label{transmap}
\rho \rightarrow 
Z^{-1} 
T^{\dagger}\Bigl( {\rm tr}_1(O \rho O^{\dagger})\otimes\openone_{l_0+1}\Bigr) T.
\ee
The number of iterations required for convergence appears to increase roughly
linearly with $l_0$ and
$\beta$, but the computational effort grows only linearly in the number of
iterations.
The locality
of the operators $\eta,O$ justifies tracing out the first site
in step (B).  After $n_{it}$ iterations, $\rho$ is approximately proportional
to the reduced density matrix
${\rm tr}_{1...n_{it}}(\exp(-\beta h^{(l_0-1+n_{it})}))$, where the partial
trace is over sites $1...n_{it}$.

To compute correlation functions of operators such as $S^z_i S^z_{i+j}$ we
follow the following procedure.  We go through steps $(1),(2),(3)$
as above, with $n_{it}$ iterations in step $(3)$.  On the last of
the $n_{it}$ iterations, after step (A) we copy $\rho$ to a new matrix
 $\rho_{corr}$.  We then replace $\rho_{corr}$ with $S^z_0 \rho_{corr}$,
thus inserting the first of the two operators.
On steps (B,C) of the last iteration we replace
of the map $\rho \rightarrow Z^{-1} {\rm tr}_1(\rho)\otimes\openone$ and
$\rho_{corr} \rightarrow Z^{-1} {\rm tr}_1(\rho_{corr})\otimes\openone$,
where the same $Z$ is used for both $\rho$ and $\rho_{corr}$.  When
then proceed through $j$ more iterations of steps (A),(B),(C), and on
the last iteration after step (A) we replace $\rho_{corr}$ with 
$S^z_0 \rho_{corr}$, inserting the second of the two operators,
before proceeding with steps (B),(C).  We then
proceed through several more iterations in which at each step we map
$\rho \rightarrow Z^{-1} {\rm tr}_1(O \rho O^{\dagger})\otimes\openone$,
$\rho_{corr} \rightarrow Z^{-1} {\rm tr}_1(O \rho_{corr} O^{\dagger})\otimes\openone$
and finally output the ratio ${\rm tr}(\rho_{corr})/{\rm tr}(\rho)$.

To compute the matrix $O$, we approximate by dividing the integral
over $s'$ into $n_{slice}$ different slices:
\begin{eqnarray}
\label{Oapprox}
O\approx 
\exp(\frac{\eta_{s(n_{slice})}}{n_{slice}}) ... 
\exp(\frac{\eta_{s(2)}}{n_{slice}}) 
\exp(\frac{\eta_{s(1)}}{n_{slice}}),
\end{eqnarray}
where $s(m)=(m-1/2)/n_{slice}$.
To compute the matrix exponential for each slice, $\exp(\eta_{s'})$ for
$s'=(1/2)(1/n_{slice}), (3/2)(1/n_{slice}), (5/2)(1/n_{slice}), ...$
we used a Taylor
series method, while to compute $\eta_{s'}$ itself we diagonalize 
$H_{s'}$ and transform $A$ into a basis of eigenvector of $H_{s'}$.
We use the fact that $H$ and $A$ conserve total
$S^z$ to speed up both this diagonalization and 
the multiplication $\rho\rightarrow O\rho O^{\dagger}$.

Since $\eta_s$ is non-Hermitian, but $||\eta_s||$ is not too large,
the Taylor series method is a good choice for computing the matrix exponential.
Eq.~(\ref{Oapprox}) approximates the $s'$-dependence of $\eta_{s'}$ by
taking its value halfway through the slice, which gives us
first order accuracy in $\partial_{s'} \eta_{s'}$ for free.
Further, $\eta_{s'}$ is in fact only very weakly dependent
on $s'$ and so a small $n_{slice}$ suffices, as seen by the following test:
set $l_0=3$.  Then, after one iteration of steps (A),(B) 
and before normalizing in step (C), the matrix $\rho$ should be equal to, 
the thermal density matrix for a three site chain.
At $\beta=1$, the largest eigenvalue should be equal to $e$.  For
a calculation with $n_{slice}=1$, the largest eigenvalue was found to
be $2.718224$; for $n_{slice}=2$, we find $2.718267$, and for $n_{slice}=3$
we find $2.718275$.

We have tested QBP by computing the susceptibility,
$\beta \sum_{j} \langle S^z_i S^z_{i+j} \rangle$, at the susceptibility peak,
using the known location\cite{betheansatz} of the peak at
$T_{max}=1/\beta=0.64085103085$.  The exact result for the susceptibility is
\be
\chi_{exact}(T_{max})=0.146926279...
\ee
while a calculation using $l_0=5$, $n_{it}=20$, $n_{slice}=20$, and correlations
up to $j=20$
gives
\be
\chi_{QBP}(T_{max})=0.146927031...
\ee
for a relative error 
of $\approx 5*10^{-6}$.
The calculation took $\lesssim 0.08$ seconds on a
1.5 GHz PowerPC G4 processor, and
using conservation of $S^z$ the largest matrix diagonalized
was $10$-by-$10$.
A larger calculation, with $l_0=9$, $n_{slice}=50$, $n_{it}=30$ improves this
to
$\chi_{QBP}(T_{max})=0.146926251...$
for a relative error of $\approx 2*10^{-7}$.

We calculated the specific heat $C$ as a function of temperature
in two ways: first, by calculating $\beta^2 \partial^2_{\beta} \ln(Z)$
using $\ln(Z)$ from the algorithm,
and, second, by calculating
$-3\beta^2 \partial_\beta \langle S_i^z S_{i+1}^z \rangle$.
The results are shown in Fig.~1, where to take derivatives we calculated
$\ln(Z)$ and $\langle S_i^z S_{i+1}^z \rangle$ for 
$\beta=0.1,0.2,0.3,...,10.0$.
As $l_0$ gets larger, the curves remain accurate to lower temperature.
The peak specific heat for $l_0=7$ was $0.349914...$ from the second derivative
calculation and $0.349717...$ from the first derivative, both of which compare
very well with the Bethe ansatz result of $0.349712...$.

The accuracy can be improved by going to larger $l_0$.  Another improvement
is to take
$H=h^{(l_0-2)}+(1/2) h_{l_0-2,l_0-1}$ and
$A=(1/2) (h_{l_0-2,l_0-1}+h_{l_0-1,l_0})$ instead of $H=h^{(l_0-1)}$ and
$A=h_{l_0-1,l_0}$.  We still have $H+A=T^{\dagger} H T + h_{0,1}$ in this
case, but the slightly different form of the perturbation seems to work
better.  The figure inset shows a comparison of Bethe ansatz data to
$l_0=9$ (where the largest matrix diagonalized is 126 dimensional).

\begin{figure}[tb]
\centerline{
\includegraphics[scale=0.3]{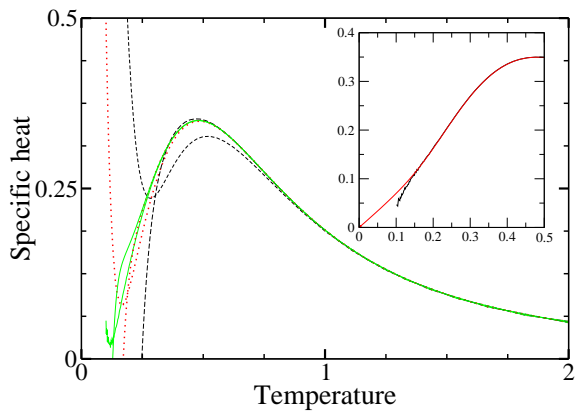}}
\caption{Specific heat against temperature for $l_0=$ 3 (dashed line),5
(dotted line), 7 (solid line).  Curves
that go negative are from
$-3\beta^2\partial_\beta \langle S_i^z S_{i+1}^z \rangle$,
while those that diverge positively
are from $\beta^2 \partial^2_{\beta} \ln(Z)$.  Inset: $l_0=9$ and
Bethe ansatz.}
\end{figure}

{\it Discussion---}
The implementation of the QBP equations here must be considered as
preliminary.  More work is needed to optimize the algorithm
and, most importantly, to quantify the sources of error.
Despite this, the method yields accurate results, giving
qualitatively correct behavior even at $l_0=3$
where QBP can be implemented as a ``pen-and-paper" technique.

In contrast to QBP, thermodynamic DMRG\cite{dmrgthermal}
computes low-lying eigenvalues for finite
size chains up to size $N=30$;
the results were accurate to quite low temperatures, but required
an extrapolation to avoid finite size effects.  In that work, chains
of size up to $N=14$ were exactly diagonalized, so if we improve the
linear algebra routines used in our implementation it should certainly
be possible to do QBP with an $l_0\sim 14$.
It may be possible to do QBP with
$l_0 \sim 30$ using DMRG techniques to avoid keeping {\it all} of the
eigenstates of $H$ and instead truncate to some smaller subset of
eigenstates of $H$.
Extrapolation of our
results suggests that this should permit access to temperatures
$T/J\sim 1/36$.

Transfer matrix DMRG\cite{tdmrg} is related to QBP
procedure in that both procedures look for the largest eigenvalue of
a transfer matrix.  In the case of transfer matrix DMRG, the transfer
matrix comes from a Trotter approximation.  In QBP,
the transfer matrix is given by Eq.~(\ref{transmap}).
Accurate results were obtained at much lower temperatures than here
in \cite{tdmrg}, but
significantly larger matrices were
diagonalized in that study,
and the higher temperature results for peak susceptibility
and specific heat do not appear to be as accurate.
It is likely that the higher accuracy of our method at high temperature
comes from the lack of Trotter error: the error becomes exponentially small
in $l_0$ once $l_0$ becomes of order $J \beta$.
We now describe a procedure that combines some of the ideas of QBP
and transfer matrix DMRG.  
Introduce $M$ copies of the system, each with density matrix
$\exp(-\beta h^{(N)}/M)$, so that the joint density matrix is
$\exp(-\beta h^{(N)}/M)\otimes ... \otimes \exp(-\beta h^{(N)}/M)$.
Let $P_i$ be the operator
that cyclically permutes the value of the spin on site $i$ between the
$M$ different copies.
Then,
${\rm tr}(P_1 P_2 ... P_N 
\exp(-\beta h^{(N)}/M)\otimes ... \otimes \exp(-\beta h^{(N)}/M))=
{\rm tr}\exp(-\beta h^{(N)})$.  Let $\rho$ be a $2^{l_0 M}$ dimensional
matrix.  
We will define a QBP procedure such that
after $n_{it}$ iterations, $\rho$ is approximately proportional
to the reduced density matrix
${\rm tr}_{1...n_{it}}(P_1 ... P_{n_{it}}
\exp(-\beta h^{(l_0-1+n_{it})}/M) \otimes ... \otimes
\exp(-\beta h^{(l_0-1+n_{it})}/M))$.
Define $O$ by Eq.~(\ref{odefn}), for $H=h^{(l_0-1)}$ at
inverse temperature $\beta/M$, and
map
\begin{eqnarray}
&\rho \rightarrow
Z^{-1} 
T^{\dagger}
\Bigl(& {\rm tr}_1(P_1 (O \otimes ... \otimes O) \rho (O^{\dagger}\otimes ...
\otimes O^{\dagger}) )\otimes \nonumber \\ &&
\openone_{l_0+1} \otimes ... \otimes
\openone_{l_0+1} \Bigr) T.
\end{eqnarray}
For $M=1$, this reduces to the QBP implementation described here, while
for $l_0=2$, this becomes very similar to the transfer matrix used
in transfer matrix DMRG.  The question is whether for $M>1,l_0>2$ more
accurate results can be obtained, possibly using DMRG to find
the fixed point $\rho$ of this transfer matrix.

QBP can be directly applied to finite size chains and to infinite or finite
trees.  The ability to compute real-space
correlation functions and handle
translationally non-invariant systems are advantages of this
method, and in future this method will be applied to disordered
systems where TDMRG will have problems.  Probably the
most interesting question is the possible application of QBP to
higher dimensional systems, by replacing the
higher dimensional lattice with a Cayley tree or Husimi cactus
with the correct local structure.

{\it Acknowledgments---}  I thank M. Chertkov for introducing me to
classical belief propagation and A. Kl\"{u}mper for supplying the Bethe
ansatz data.
This work supported by U. S. DOE Contract No. DE-AC52-06NA25396.


\begin{thebibliography}{99}
\bibitem{locEXP} For small enough $\beta$, the operator
$\int {\rm d}\omega \frac{\exp(\beta \omega)-1}{\omega}
A^{\omega,s}$ is local as may be shown by a power series expansion, but
at larger $\beta$ this fails.
\end{thebibliography}
\end{document}